# Relativistic velocity addition law derived from a machine gun analogy and time dilation only

**Bernhard Rothenstein[1] and Stefan Popescu[2]**
1) Politehnica University of Timisoara, Physics Department,
Timisoara, Romania brothenstein@gmail.com
2) Siemens AG, Erlangen, Germany stefan.popescu@siemens.com

***Abstract.*** *We consider a scenario that involves a machine gun, the bullets it fires and a moving target, considered from the rest frame of the machine gun and from the rest frame of the target respectively. Involving the special relativity via its two postulates and the time dilation formula we derive the relativistic velocity addition law showing that it leads to the Lorentz transformations for the space-time coordinates of the same event.*

## 1. Deriving the relativistic velocity addition law without using the Lorentz transformations

Many derivations of the relativistic addition law of parallel velocities without use of the Lorentz Transformations (LT) are known.[1,2,3,4,5] Some of them are based on thought experiments that require knowledge of the time dilation and the length contraction effects[1,4,5]. Other derivations involve the Doppler Effect in the optic domain considered from three inertial reference frames in relative motion.[6] A few derivations simply involve only the principle of constancy of the light velocity[2]. Such derivations are interesting for the teaching of special relativity theory since the relativistic addition of velocities leads directly to the LT.[7]

The derivation we propose is based on a machine gun - target analogy[8] of the acoustic Doppler Effect, considered from the rest frame of the machine gun and from the rest frame of the target. This derivation involves special relativity via the formula that accounts for the time dilation effect which in turn is a direct consequence of the two relativistic postulates as it is best shown by the thought experiment that involves two light clocks in relative motion.[9]

The scenario we follow involves a stationary machine gun (MG) located at the origin O of its rest frame I(XOY) and a target T located at $x=x'=0$ when $t=t'=0$ and moving with constant velocity V in the positive direction of the OX axis[9]. The machine gun fires successive bullets at constant time intervals $\Delta t_e = t_e - 0$ measured as a difference between the readings of a clock $C_0$ located in front of the MG, which by definition represents a proper time interval. The bullets move with constant velocity U in the positive direction of the OX axis and hit the moving target T at constant time intervals $\Delta t_r = t_r - 0$. The time interval $\Delta t_r$ is measured as a difference between the readings of a clock $C$ located where the bullet hits the target and the reading of clock $C_0$ when the target is located in front of it. The clocks $C$ and $C_0$ are synchronized in accordance with the procedure proposed by Einstein. By definition $t_r$ represents a coordinate time interval.

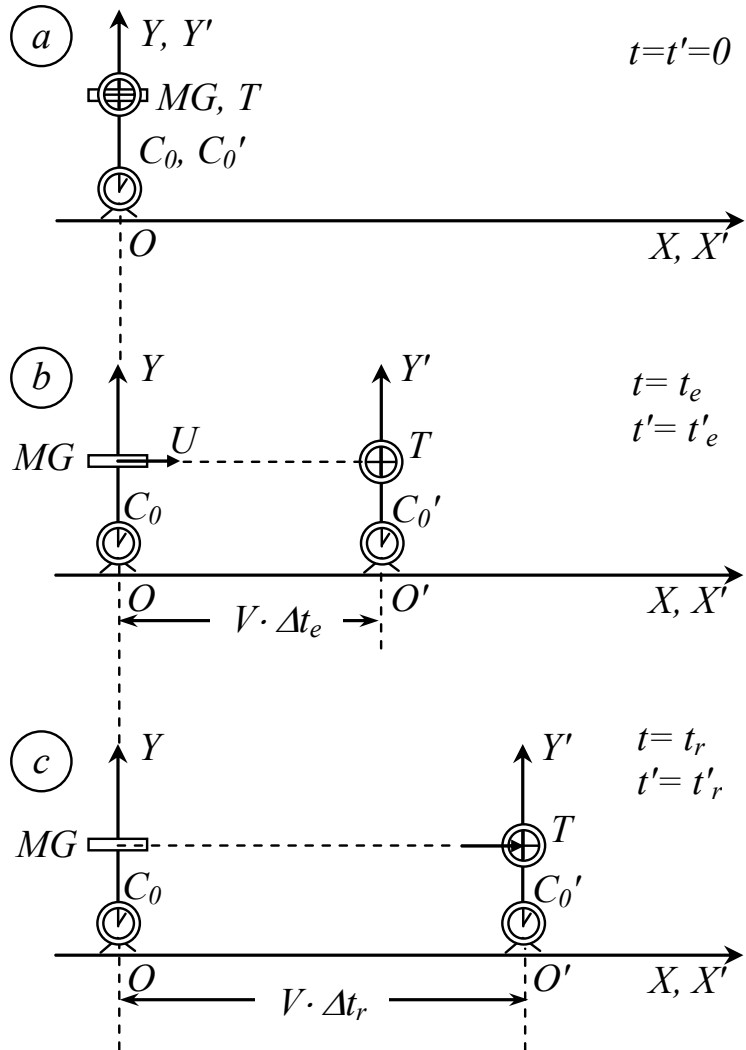

**Figure 1a.** The machine gun MG and the target T are located at the same point in space the clocks the clocks $C_0$ and $C'_0$ attached to them reading both a zero time.
**Figure 1b.** Position of the target T when the machine gun fires a bullet.
**Figure 1c.** Position of the target T when the bullet hits it.

Equating the distance traveled by the second fired bullet with the distance traveled by the target between the reception of the second and of the first bullet, we obtain (U<V)

$$Vt_r = U(t_r - t_e) \tag{1}$$

from which we obtain

$$t_r = \frac{t_e}{1 - \dfrac{V}{U}}. \tag{2}$$

which holds for the corresponding time intervals $\Delta t_e$ and $\Delta t_r$ respectively.

Let $C'_0$ be a clock attached to the target. It reads $t'=0$ when it is located in front of the machine gun but reading $t'_r$ when the second hits it, where by definition $\Delta t'_r = (t'_r - 0)$ represents a proper time interval related to $\Delta t_r$ by the time dilation formula

$$\Delta t_r = \frac{\Delta t'_r}{\sqrt{1 - \dfrac{V^2}{c^2}}} \tag{3}$$

With (2) and reformatting (3) we obtain

$$\Delta t'_r = \Delta t_e \frac{\sqrt{1 - \dfrac{V^2}{c^2}}}{1 - \dfrac{V}{U}}. \tag{4}$$

which relates the proper time interval $\Delta t_e$ at which the machine gun fires successive bullets measured in its rest frame and the proper time interval $\Delta t'_r$ at which the successive bullets hit the target measured in its rest frame.

Let I'(X'O'Y') be the rest frame of the target. There the machine gun moves with velocity V in the negative direction of the axes O'X', whereas the bullets move with velocity U' in the positive direction. We equate in this case the distance traveled by the machine gun in the time interval $\Delta t_e$ and the distance traveled by the bullet i.e.

$$Vt'_e = U'(t'_r - t'_e) \tag{5}$$

from which we obtain expressed as a function of time intervals

$$\Delta t'_e = \frac{\Delta t'_r}{1 + \dfrac{V}{U'}}. \tag{6}$$

In this case $\Delta t'_e$ represents a time interval measured as a difference between the readings of a clock C' of the reference frame I' located where the bullet is fired ($t'_e$) and the reading $t' = 0$ of the clock $C'_0$ attached to the target. The two clocks mentioned above are synchronized in the reference frame I' following the same synchronization procedure as in I. The clock commoving with the MG measures a proper time interval $t_e$ between the emissions of two successive bullets, which is related to $t'_e$ by the time dilation formula

$$t_e = \frac{t'_e}{\sqrt{1-\frac{V^2}{c^2}}} \qquad (7)$$

with which (6) becomes expressed as a function of time intervals

$$\Delta t_e = \Delta t'_r \frac{\sqrt{1-\frac{V^2}{c^2}}}{1+\frac{V}{U'}}. \qquad (8)$$

Eliminating $t_e$ and $t'_r$ between (4) and (8) we obtain that the velocities of the **same** bullet relative to I and respectively to I' are related by

$$U = \frac{U'+V}{1+\frac{U'V}{c^2}} \qquad (9)$$

or by

$$U' = \frac{U-V}{1-\frac{UV}{c^2}} \qquad (10)$$

In this way we have derived the relativistic addition law of parallel velocities without prior knowledge of the Lorentz transformations.

**2. Lorentz transformations from the relativistic addition law of velocities.**

Consider the moving bullet from the stationary reference frame I of the machine gun. After a given time of motion the bullet hits a target and generates the event E(x=Ut,y=0,t=x/U) whereas, when detected from I' it generates the event E'(x'=U't',y'=0,t'=x'/U'). The motion of the bullet is described in I by

$$x = Ut \qquad (11)$$

and in I' by

$$x' = U't' \qquad (12)$$

respecting the initial conditions (t=t'=0, x=x'=0). The scenario we follow ensures the fact that the two events take place at the same point in space when the clocks C(x=Ut,y=0) and C'(x'=U't',y'=0) located at that point read t and respectively t' or, as relativists would say, they represent the **same** event.

Combining (11) and (12) we obtain

$$\frac{x}{t} = \frac{x'}{t'} \frac{1+\dfrac{V}{U'}}{1+\dfrac{VU'}{c^2}} \qquad (13)$$

Equation (13) suggests considering that

$$x = f(V)x'(1+\frac{V}{U'}) \qquad (14)$$

and

$$t = f(V)t'(1+\frac{VU'}{c^2}) \qquad (15)$$

where f(V) represents an unknown function of the relative velocity but not of the space-time coordinates of the events involved. This guess is based on the logical evidence that the transformation equations (14) and (15) should be linear, in order to transform the uniform motion of the bullet in one of the reference frames into a uniform motion in the second frame. We obtain the algebraic structure of f(V) by imposing the condition that for U'=0 (15) accounts for the time dilation i.e.

$$f(V) = \frac{1}{\sqrt{1-\dfrac{V^2}{c^2}}} \qquad (16)$$

and so (14) and (15) become

$$x = x'\frac{1+V/U'}{\sqrt{1-V^2/c^2}} = \frac{x'+Vt'}{\sqrt{1-V^2/c^2}} \qquad (17)$$

$$t = t'\frac{1+VU'/c^2}{\sqrt{1-V^2/c^2}} = \frac{t'+Vx'/c^2}{\sqrt{1-V^2/c^2}} \qquad (18)$$

Thus we have derived the Lorentz transformations for the space time coordinates of an event generated by the same bullet as detected from two inertial reference frames in relative motion taking place at the same point in space when the synchronized clocks of the two frames located at that point read t and t' respectively. Under such conditions relativists say that the Lorentz transformation relates the space-time coordinates of the same event.

**Conclusions**

A simple and transparent scenario considered from two inertial reference frames in relative motion and special relativity involved via its two postulates and time dilation leads to the relativistic addition law of parallel velocities which leads at its turn in a transparent way to the Lorentz transformations for the space-time coordinates of the same event. We found a direct and simple inference way that starts with the two relativistic postulates and proceeds via time dilation formula and the deduction of the relativistic velocity addition law to finally reveal another deduction for the Lorenz transformations.